# Prediction is very hard, especially about conversion*

Predicting user purchases from clickstream data in fashion e-commerce

Luca Bigon
Tooso Labs
San Francisco, CA USA
luca.bigon@tooso.ai

Giovanni Cassani
CLiPS, Department of Linguistics
University of Antwerp
Antwerp, Belgium
giovanni.cassani@uantwerpen.be

Ciro Greco, PhD
Tooso Labs
San Francisco, CA USA
ciro.greco@tooso.ai

Lucas Lacasa, PhD
School of Mathematical Sciences
Queen Mary University
London, United Kingdom
l.lacasa@qmul.ac.uk

Mattia Pavoni
Tooso Labs
San Francisco, CA USA
mattia.pavoni@tooso.ai

Andrea Polonioli, PhD
Tooso Labs
San Francisco, CA USA
andrea.polonioli@tooso.ai

Jacopo Tagliabue, PhD[†]
Tooso Labs
San Francisco, CA USA
jacopo.tagliabue@tooso.ai

## ABSTRACT

Knowing if a user is a "buyer" vs "window shopper" solely based on clickstream data is of crucial importance for e-commerce platforms seeking to implement real-time accurate NBA ("next best action") policies. However, due to the low frequency of conversion events and the noisiness of browsing data, classifying user sessions is very challenging.
In this paper, we address the clickstream classification problem in the fashion industry and present three major contributions to the burgeoning field of A.I.-in-fashion: first, we collected, normalized and prepared a novel dataset of live shopping sessions from a major European e-commerce fashion website; second, we use the dataset to test in a controlled environment strong baselines and SOTA models from the literature; finally, we propose a new discriminative neural model that outperforms neural architectures recently proposed by [1] at Rakuten labs.

## CCS CONCEPTS

• Computing methodologies → Neural networks; • Applied computing → Online shopping

## KEYWORDS

Clickstream prediction, intent detection, time-series classification, deep neural network.

## 1 Introduction

The extraordinary growth of online retail in recent years [2] had a significant impact on the fashion industry, where almost 25% of all fashion-related transactions are now happening online [3]. In a vertical where new digital players keep emerging, A.I. is becoming a key differentiator between companies which are thriving in this Big Data world and more traditional organizations which are left behind [4].
While conversion rate (the proportion of user sessions ending with a monetary transaction vs the total number of sessions) fluctuates across countries, industries and websites [5] the vast majority of sessions are from users with weak buying intention ("window shoppers"). Being able to turn window shoppers into converting customers is a key priority for fashion brands, as small increases in e-commerce conversion rates can translate into massive revenue uplifts for businesses. Moreover, while not all online visitors can obviously be turned into online conversions, in today's omnichannel landscape webrooming is on the rise and engagement with users can lead website visitors to complete their purchases in physical stores [6]. The availability of increasingly smarter technological solutions has recently resulted in fashion brands becoming interested in next-best-action marketing and personalization to achieve tech-gagement. Specifically, if a website knew in advance the "true" intention of its users, it could leverage the possibilities offered by the digital world to offer a real-time personalized experience and cater the right messages and offers to the right customers, which is known to increase revenues [7]. Identifying users which are very likely to convert is therefore both potentially very valuable and very hard, due to class imbalance and the general noisiness of browsing data. In *this* paper, we present the ongoing research that Tooso Labs is conducting on real-time intent detection, by marrying Artificial Intelligence and deep domain knowledge over proprietary fashion data. While this is submitted as a "work in progress" paper, we are already in a position to make three substantial contributions:

1) we introduce a new dataset of session data from a major European e-commerce fashion player: we plan to announce the release of the full dataset under a research friendly license as part of KDD participation;
2) we make the first (to our knowledge) thorough benchmark of several baselines and more advanced models over a standard, reproducible dataset;
3) we improve over known models with a new neural architecture; moreover, we sketch in the final section approaches from the physics of non-linear system as an active area of research in our lab.

The paper is organized as follows: in Section 2 we precisely define the intent detection problem and introduce our new dataset, describing the steps we took to transform the raw data into the final dataset; in Section 3 we detail our methodology and describe all the models in our study. In Section 4 we present results and a preliminary analysis before concluding, in Section 5, with some final remarks and our plans for future work.

## 2. Problem Statement and Dataset

### 2.1 Problem Statement

The final industry goal of this project is to give fashion retailers a reliable way to *incrementally* estimate the probability of a given session being a "buying" vs "non-buying" session. The constraints of the problem are therefore the following:

1) input data are anonymized session data from a fashion store (e.g. clicks from a fashion e-commerce);
2) data are incrementally "streamed" into the system as the user session progresses; at any given data point, the "real" length of the session is unknown.

In *this* paper, we start with some simplifying assumptions to bootstrap our investigation into promising methods for intent detection.[1] More formally: let $e_1, e_2, \ldots e_n$ be a sequence of "click events" with associated metadata $m_1, m_2, \ldots m_n$ and timestamps $t_1, t_2, \ldots t_n$ (where $t_1 < t_2 < t_3 \ldots$), performed by user $U$ on website $W$, with each $e \in \{C \mid$ "view", "click", "detail", "add-to-cart", "remove-from-cart", "buy"$\}$. Events $e_1, e_2, \ldots e_n$ are then partitioned into non-overlapping *sessions*, i.e. sequences of events such that for any pair of consecutive events $e_k$ and $e_{k+1}$ with timestamp $t_k$ and $t_{k+1}$, $e_k$ and $e_{k+1}$ belong to the same session $s$ if and only if $(t_{k+1} - t_k) <= 1800$ seconds (30 minutes is the industry standard, as for example found in Google Analytics [8]). For each session $s$, we say that $s$ is a "buy-session" (BUY, for short) if and only if there is at least one $e$ in $s$ such that $e \in$ "buy"; we say that $s$ is a "no-buy-session" (NOBUY, for short) if and only if, for every $e$ in $s$, $e \notin$ "buy". We are now in a position to precisely state a simplified version of the intent problem, which will be used as a reference for the remainder of the paper (but see the final section for our remarks on the "original" version):

(IP) given a session $s$ from $U$ on $W$, classify $s$ as BUY or NOBUY.

Two final considerations are in order:

1. when evaluating a full session $s$, if $s$ does indeed contain a "buy" event $e$, the session is cut just before $e$ (to avoid trivializing the inference challenge);
2. while we allow for the usage of timestamps and metadata (e.g. product features) in the general formulation of the problem, the benchmarks below are based on a symbolized version of user sessions (see sections 2.3 and 3 for details on data preparation

---

[1] The task is also known in the literature as "clickstream prediction": we shall use the labels interchangeably.

and methodology): our methods are able to make reliable inferences with very limited information.

## 2.2 The Tooso Fashion Clickstream

In *this* paper, we present for the first time to the fashion and A.I. community our "Tooso fashion clickstream dataset" (TFCD), a dataset containing data from real user sessions on a major European e-commerce fashion website. Although the volume of A.I. and Machine Learning papers devoted to retail and fashion is growing [4], a lot of progress in the field is still hindered by the impossibility of reproducing published results, either because of lack of shared data or working code (or both).

As part of our participation to KDD, we plan to time the release of the dataset in a research friendly format so that it will be easy for the community to a) reproduce our results, running our code over the TFCD, b) contribute to the clickstream prediction problem by developing and sharing new and improved architectures. Our goal in the fashion and A.I. community is to accomplish for the clickstream prediction challenge what the Fashion MNIST dataset achieved for computer vision tasks [9]. TFCD is built using a subset of user events from the e-commerce of a major (>1B year turnover) fashion group in Europe: all events in the dataset are from the period 06/29/2018 al 07/18/2018.[2]

Data has been collected in compliance with the legislation through a web pixel [10] [11] placed on the target property; in particular, any click performed by final users browsing the website is collected in Tooso servers with anonymized user id, timestamp, product metadata (if available) (Tooso Analytics follows the Google Measurement Protocol [12]). Once cleaned, anonymized and sampled through the anonymized user id, events are assigned a session using Tooso's "sessionization" pipeline, i.e. events are assigned the same unique session id if they belong to the same session. The resulting raw dataset has the following features:

| Total events | 5,405,565 |
|---|---|
| Total sessions | 443,652 |
| 0/25/50/75/100th percentile for session length | 1 / 1 / 4 / 12 / 7579 |
| Event types | 6 - [view, detail, add, remove, buy, click] |
| Total unique users | 212037 |
| Total unique products (SKUs) | 33237 |

Table A: **features of the dataset.**

A typical row from the dataset looks like the following:

| Field | Type | Example |
|---|---|---|
| client hashed id (cookie-based id) | 36 char UUID | 2c1b7958-af02-4dc2-80a0-e3247fd96706 |
| user hashed id (if user is logged in) | 36 char UUID | f3cc5e93-48e2-47b1-b19e-93607cb61276 |
| session id | 36 char UUID | 1186c594-7fff-4de4-9af7-d559ce9b1f5e |
| timestamp | long (epoch timestamp in ms) | 1530480640476 |
| event hashed id | 36 char UUID | a97fb677-dab9-46d0-b3c0-265f50fcf36a |
| event type | one between [view, detail, add, remove, buy, click] | "add" |
| product hash | 36 char UUID | 62d5a348-8a78-4e11-af32-6ff51e12bd81 |
| product metadata as embeddings | JSON | { top_10_pca_factor s = [...] } |

Table B: **Typical row from the dataset**

Please note that even if all identifiers were anonymous in the first place (e.g. data collection through the web pixel), they get hashed a second time to further remove chances of re-identification, making impossible to match cookie data in browsers with data released in the dataset.[3] In the next section, we describe a simplified version of the TFCD used in *this* paper.

---

[2] While it is known that seasonality may affect behavioral patterns in fashion shopping, using sessions from a single month should not affect the reliability of the benchmark presented here, as sessions are fed as completely independent time-series to all the models.

[3] At the moment of drafting *this* paper, we are still actively working with our lawyers to understand how to effectively release product data in compliance with all regulations (e.g. anonymized PCA components of textual and image embeddings). The final details on dataset structure will be the result of the current legal discussions, so small adjustments and changes are possible. Similarly, the code will be made available to the general public under the company Github account at the time of releasing TFCD.

## 2.3 The symbolized clickstream dataset

Instead of using the full power of the TFCD, we decided to start with simplifying assumptions on the data side; by stripping some of the more specific information and using only minimal time-series features, we present results which are very interesting since i) we implicitly make our findings immediately applicable to a vast array of use cases and companies where only high-level data (not pixel fine-grained data) are available; ii), we can readily compare very different models in a principled and fair way, controlling for model-specific ability to combine features. In particular, information on users is retained only insofar as each session represents a distinct user in a 30-min window browsing on the website; otherwise the model is "memory-less" so that signals coming from previous sessions cannot be exploited for predictions (e.g. considering the amount of non-recurrent customers in fashion commerce, this feature would actually strengthen industry interest for a working model). Starting from TFCD, we "symbolize" user sessions, so that, for each $e$ in a session $s$, the only information we retrieve is the event type, represented with integers to simplify implementation.

A typical user session is therefore represented in the final dataset as:

4, e00caaf5048f482fa73a42f3f9e228966c29a585
2-1-1-2

where 4 is the session length, e00… is the session unique id, and [2, 1, 1, 2] is the symbolized clickstream (in this case, [detail, view, view, detail]). After cleaning for too long/short sessions (10 < len(s) < 200), the following table resumes the benchmark dataset main statistics:[4]

|  | ALL | BUY | NO BUY |
|---|---|---|---|
| # events | 3796088 | 328543 | 3467545 |
| # sessions | 130,572 | 7176 (5.5%) | 123,396 |
| 0/25/50/75/100th percentile for session length | 10 / 14 / 20 / 34 / 200 | 10 / 18 / 33 / 60 / 200 | 10 / 13 / 20 / 33 / 200 |
| mean (± sd) session length | 29.07 ± 25.17 | 45.78 ± 37.8 | 28.1 ± 23.88 |

Table C: **Benchmark dataset main statistics**

[4] These are data and proportions as calculated before (random) downsampling.

By excluding sessions shorter than 10 events, we discarded about 70% of the original data. By further excluding sessions longer than 200 events we lost an extra 1% of the data. Finally, ~700 BUY sessions became shorter than 10 events after cutting the session before the first buy event: these sessions were further removed, causing a further loss of 0.58% of the data, yielding a corpus consisting of 7,176 BUY sessions and 123,396 NOBUY sessions (see table C). The resulting corpus was first balanced by randomly downsampling the NOBUY sequences to yield 14,352 total sessions and then split in training (70% of the sequences from each class), validation and test (15% of the sequences each for each class).

Figure 1 shows the distribution of events in the dataset comparing BUY sequences with NOBUY sequences. While the proportions of *views* and *clicks* are fairly similar in the two types of sequences, we see that, in line with expectations, NOBUY sequences have many more *detail* events while BUY sequences have more than double the amount of *add to cart* actions. Finally, contrary to expectations, BUY sequences have proportionally more *remove from cart* events as compared to NOBUY sequences.

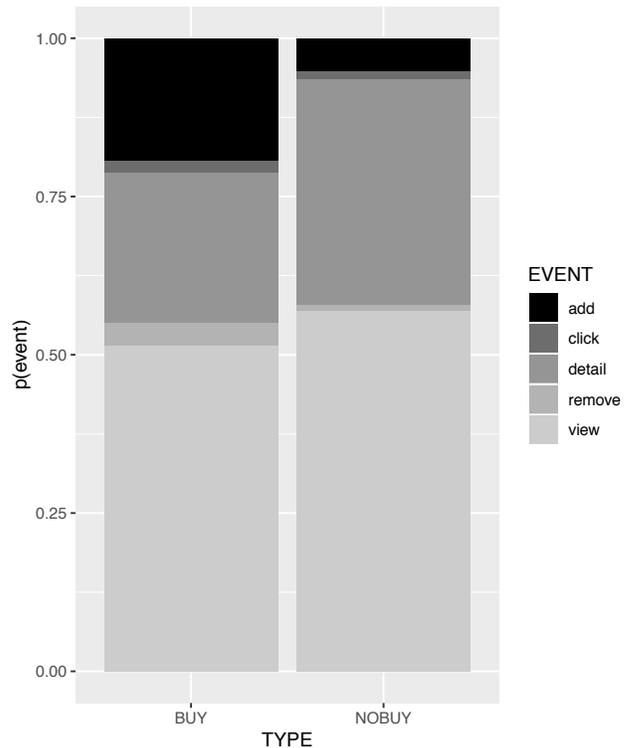

Figure 1: **distribution of events in the dataset, comparing BUY sequences with NOBUY sequences.**

Figure 2 shows the difference between state transition probabilities in BUY vs NOBUY sequences (P(S|BUY) - P(S|NOBUY)), color-coded in a red-blue scale (such that, for example, heavily red transitions are far more likely for NOBUY sessions).

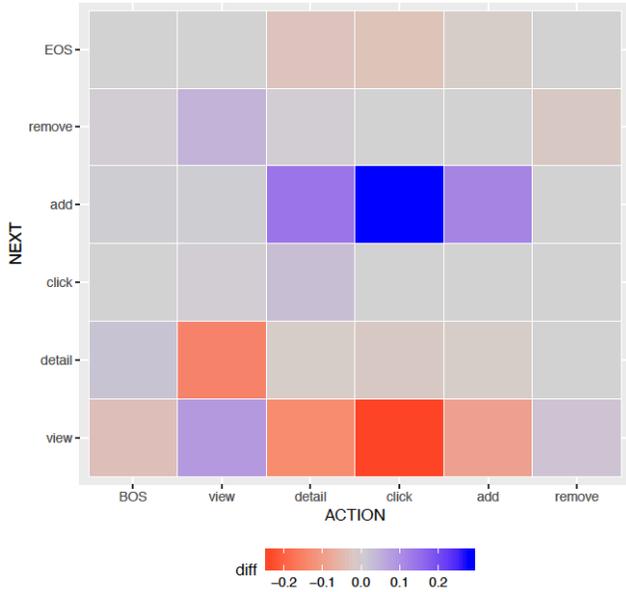

Figure 2: **difference between state transition probabilities in BUY vs NOBUY sequences.**

## 3. Methodology

In this section, we describe all the models and related implementation details of the algorithms for the analysis in Section 4. We divide our overview in three main subsections: a generic baseline, comparison with existing clickstream models, and new contributions.[5]

## 3.1 Baseline

As often recognized in the literature [1], there are strong resemblances between the clickstream prediction problem and standard topics in the NLP community, as many NLP-tasks are i) sequential, ii) based on a discrete set of atomic states [13] [14]. As our general baseline, we picked an n-gram Naive Bayes model.

Clickstreams are featurized extracting n-grams and then treated as a bag of n-grams. Normalized frequency counts for each n-gram under the BUY and NOBUY sequences are computed and used for classification. At prediction time, given a new session to be classified, we make a decision using Bayes rule: the distribution of BUY/NOBUY in the training corpus is the prior, and the probability under BUY/NOBUY sequences is the likelihood, i.e.

$$P(\text{BUY}|s) = \frac{[P(\text{BUY}) * P(s|\text{BUY})]}{[P(\text{BUY}) * P(s|\text{BUY}) + P(\text{NOBUY}) * P(s|\text{NOBUY})]}$$

Equation 1

For any session $s$, we predict the class associated with the highest probability, i.e. $s \in$ BUY iff $P(\text{BUY}|s) > P(\text{NOBUY}|s)$, $s \in$ NOBUY otherwise. We explored n-grams with n between 1 and 5, and found that penta-grams (n=5) provide the best accuracy.

### 3.2 Literature comparisons

*3.2.1 Markov chains*

We reproduce the methodology of the influential [15], which borrows from the long-standing idea that browsing activities are properly modelled as Markov processes [16] [17]. In particular, two separate Markov chains are trained for BUY/NOBUY sequences.

At prediction time, given a new session to be classified, we make a decision using the same Equation 1 above. For any session $s$, we predict the class associated with the highest probability, i.e. $s \in$ BUY iff $P(\text{BUY}|s) > P(\text{NOBUY}|s)$, $s \in$ NOBUY otherwise. We run several experiments to pick the best degree for the final chains and found that chains of order 5 provide the most reliable classification accuracy (in line with [1]). Hence, the probability of each state given the preceding penta-gram is considered. However, the increase with respect to chains of order 4 is minimal, suggesting that increasing the complexity of the chain further will not provide sizeable advantages.

*3.2.2 LSTM language model*

A recent paper by the technology team of a leading Japanese e-commerce website [1] reported improvements over the MC approach in [15] using LSTMs [1].

While they frame the problem as a three-fold classification (*purchase*, *abandon* or *browsing-only*), they used the same idea of "mixture models" as in [15] just replacing MCs with probabilities from a neural network model (token probabilities are read off intermediate softmax layers in each LSTM model). We built two LSTMs (BUY, NOBUY) with as many input units as there are input classes (plus a BOS and an EOS symbol) - one-hot encoded - and the same number of output units. We used Cross Entropy as our loss function and trained the network using Adam. In line with [1], we considered architectures with 1 hidden layer and 4 different values for the number of hidden units (10, 20, 40, 80). Hence, in between the input and output layers, we included one LSTM layer of dimensionality (input units x hidden units) and one fully connected layer with dimensionality (hidden units x output units).

---

[5] For the sake of brevity, we will not get into specific architectural details on the actual implementation and deployment choices; as a general overview, baseline models were manually implemented in Python, DL models were implemented with PyTorch and the data pipeline and model serving infrastructure was built using AWS PaaS services (s3 for storage, EMR/Spark for data extraction and minimal pre-processing, SageMaker for DL training and testing).

We also explored different values for the learning rate (0.01, 0.001) and for the batch size (10, 20, 50). Unlike [1], we did not consider models with 2 hidden layers and did not explore the effect of dropout.

We trained each model with early stopping on the accuracy on the validation set, with a patience of 10 and a maximum number of epochs of 50. At prediction time, each sequence is passed through both LSTMs and the probability of every state in the sequence is retrieved from the softmax layer. Classification happens in the same way as in the two previous models (see EQUATION 1 above).

The grid search revealed that models with different parameterizations tend to perform similarly, showing that room for optimization is limited, at least within the current architecture. However, a larger hidden layer, a smaller learning rate and a large batch size provided slightly but consistently better performance. Hence, we tested an architecture with 80 hidden units, trained with a learning rate of 0.001 and a batch size of 50. Accuracy on the validation set tended to peak in between 30 and 45 epochs, then plateaued and started to wiggle.

### 3.3 Novel contributions

*3.3.1 Seq2Label*

Building on the LSTM model in 3.2.2, we implemented a discriminative classifier as an alternative way to conceptualize the clickstream problem. As usual, training happens by feeding batches of sequences and training the model to minimize prediction error from the sequence to the binary output class. This architecture consisted of one LSTM layer of dimensionality (input units x hidden units) and one fully connected layer with dimensionality (hidden units x 1), whose output was transformed using the sigmoid activation function before computing the loss.

Two pooling strategies were explored, changing the information that is used to classify sequences: taking the output of the LSTM at the last time step (ignoring padding indices) and taking the average LSTM output over the entire sequence (again ignoring padding indices). The pooled output of the LSTM was then passed through the fully connected layer and transformed using the sigmoid activation function, which was then taken as the prediction given the sequence. Binary Cross Entropy Loss is used to quantify the error and back-propagate it (Adam was again the chosen optimizer). We tested the same hyper-parameter settings as in the LSTM language model and trained each model with early stopping, considering accuracy on the validation set as the target variable with a patience of 10.

Again, performance was stable across different parameter constellations. We picked one parametrization for each pooling heuristic. Architectures trained with average pooling tended to perform better when they had more hidden units, were trained with a higher learning rate and using large batches (hence we tested an architecture with 80 hidden units, a learning rate of 0.01 and a batch size of 50). Architectures trained considering the LSTM output at the last time step were better when smaller and trained on small batches. Hence, the final model we tested consisted of 20 hidden units and was trained with a batch size of 10 and a learning rate of 0.01. Training was faster as compared to the Generative LSTM, particularly when classification was optimized based on the last LSTM output: in this case, validation accuracy peaked before 10 epochs. Architectures with average pooling trained for a bit longer before stopping, typically between 10 and 20 epochs.

## 4 Results and analysis

After deciding on the best parameter choices for our models, we tested them on the test split of the corpus. However, since LSTMs depend on random initialization, we trained 10 different instances of the same model in order to make sure that differences in performance did not depend on stochastic processes. Naive Bayes and Markov Chain classifiers were not trained multiple times, since there is no stochastic process involved. In the following table we report average accuracy scores on the test set, and provide the standard deviation over 10 runs in parentheses when necessary.

| Model | Accuracy |
| --- | --- |
| Naive Bayes | 0.821 |
| Markov Chain | 0.882 |
| LSTM - Language Model | 0.909 (± 0.004) |
| LSTM - S2L ('avg' pooling) | 0.927 (± 0.003) |
| **LSTM - S2L ('last')** | **0.932 (± 0.002)** |

Table D: **Average accuracy scores**

Some comments are in order:

1. The Naive Bayes classifiers does unsurprisingly worst since it does not capture any time-dependent information.
2. The Markov Chain classifier improves substantially, showing that the information captured by state transitions is important to the task.

3. The Generative LSTM, which replaces the language model based on the Markov Chain transition matrix with LSTMs, performs better than its direct counterpart, in line with [1].
4. Both S2L models outperform the generative models, with the architecture using the last LSTM output outperforming its counterpart (the difference is statistically reliable at the 99% confidence level).

Further differences between the two S2L architectures concern training speed and complexity of the network: the architecture which does average pooling fares worse under both regards, taking longer to train and having to learn more parameters due to a larger hidden layer. However, both complexity and training time still improve with respect to the generative LSTM.

The fact that the S2L model which considers the LSTM output at the last time step learns so well and so fast could indicate that right before completing a purchase, users tend to have very distinctive clicks (although by-class precision and recall are in line with the global accuracy, suggesting that the model does not outperform its competitors because it does one thing much better than the others). Still, if this is true, then this S2L architecture could break down when classifying incomplete sequences which are still unfolding. Its counterpart which implements average pooling may therefore have an advantage for early prediction, despite lagging behind in whole sequence categorization accuracy.

Another advantage of the S2L discriminative LSTMs is that they scale naturally to any number of classes, without any need to train a different model on sequences from each class and then perform MAP classification, as in Equation 1. The disadvantage, however, is that at every new event the whole sequence has to be fed again to the classifier since there is no knowledge of the generative process which likely yields the sequence as it unfolds.

## 5 Conclusions and directions for future work

In this paper we presented preliminary but encouraging results in the clickstream prediction challenge for the fashion industry. Moreover, we collected and curated a new dataset for the A.I. community to serve as a standard, public benchmark for this type of tasks, in the same spirit as the Fashion MNIST dataset [9] was curated for computer vision in fashion.

Several extensions of course are being considered. First and foremost, we plan to adapt and recast all the featured architectures in the "original" prediction setting (Section 2.1), where classification is made incrementally as new data points arrive (some architectures, as briefly discussed already, are easier to "port" than others); a key challenge would be to understand how to train each model so that it is not biased by the high prior for NOBUY.

Second, leveraging a standard dataset as a stable and reliable benchmark, new algorithms can be implemented and rigorously tested: in this respect, our current experiments are focusing on visibility graphs (a time-series analysis tool from the physics of complex systems [18] [19]), autoencoders (i.e. treating imbalanced classification as a problem in anomaly detection), and a Multi-Task objective LSTM (which *jointly* optimizes the prediction of the next state *and* the classification of the sequence).

Finally, most of the architectures presented can be adapted to add further metadata to the barebone symbolized sessions that have been used in the current study. While inter-click dwell times are the low-hanging fruit here [1], the most interesting results may well come by adding metadata (and metadata-derived features) from the fashion products the users interacted with in their session.

All in all, whatever architecture and feature subset will be proven to the best at this particular task, we *do* believe that a commitment to open science and (whenever possible) open data is the key to foster the growth of reliable and reproducible progress in the A.I.-in-fashion community.

## STATEMENT OF AUTHOR CONTRIBUTION

Due to the nature of the task, the real-time nature of the use case, the scale of web data, the business and legal challenges involved in the treatment of sensitive industry data, building an end-to-end clickstream prediction pipeline requires several skills. All seven authors contributed equally to the final result, each according to their own primary expertise.

## ACKNOWLEDGMENTS

The authors are very grateful for the help of the editors and reviewers in shaping the final version of the article. The authors also wish to thank Tooso Inc. for providing the computational infrastructure and funding for the project. Jacopo Tagliabue wishes to thank Tooso's clients too, as they have been instrumental in the success of the company and have been always very receptive to the possibilities opened by A.I. in retail and fashion.

.